\documentclass[12pt]{article}
\usepackage{graphics,amsmath }
\begin{document}

{\bf \large Intermittency and Localization}

\bigskip
G. Yaari,$^{1,2}$ D. Stauffer$^{3,2}$ and S. Solomon$^{2,1}$

\bigskip  $^1$ {\small Institute for Scientific Interchange, via S. Severo 65, I-10113 Turin, Italy}

\bigskip $^2$ {\small Racah Institute of Physics, Hebrew University, IL-91904 Jerusalem, Israel}

\bigskip  $^3$ {\small Institute for Theoretical Physics, Cologne University, D-50923 K\"oln, Germany}

\bigskip

{\bf Article Outline}
\renewcommand{\labelenumi}{\bf{\theenumi.}}
\renewcommand{\labelenumii}{\bf{\theenumi.\arabic{enumii}.}}
\renewcommand{\labelenumiii}{\bf{\theenumi.\arabic{enumii}.\arabic{enumiii}.}}

\begin{enumerate}
 \item Definition
\item Introduction 
 \item Logistic systems: From Malthus until today.
\begin{enumerate}
\item "auto-catalysis".
\item Real-Life examples.
\begin{enumerate}
\item The A(utocatalysis)-Bomb.
\item The B-Bomb: autocatalysis and localization in immunology.
\item The Tulip Bomb.
\end{enumerate}
\item Extensions of the classical logistic system.
\item The danger of being mean - Simple examples.
\end{enumerate}
 \item Minimal extensions to the classical logistic system.
\begin{enumerate}
\item Case 1: The Generalized Lotka-Volterra System. 
\item Case 2: The ”AB Model”
\end{enumerate}
 \item Applying these models to real-life systems.
 \item Future Directions
\end{enumerate}
\bigskip
\newpage
{\bf \large Glossary}
\begin{itemize}
\item  \textbf{Auto-catalysis.} In systems that go through several reactions, the reaction is called \textbf{autocatalytic} if the reaction product is itself the catalyst for that reaction.
\item {\bf Exponential growth.} 
An autocatalytic reaction is usually described with a simple linear, first order differential equation. The solution for it is an exponential increasing/decreasing function.
\item {\bf Logistic growth.} 
If one adds a saturation term (of power two) to the linear first order differential equation which describes exponential growth, the resulting solution saturates instead of ever-lasting grow. The solution to this system is described with a \textbf{logistic curve} and the system is said to follow a \textbf{logistic growth}.
\item {\bf Reaction-diffusion systems.} 
Reaction-diffusion systems are mathematical models that describe how the concentration of one or more substances distributed in space changes under the influence of two processes: local reactions in which the substances are converted into each other, and diffusion which causes the substances to spread out in space.
\end{itemize}

{\textbf{Acknowledgements}}
The present research was partially supported by the STREPs CO3 and DAPHNet of EC FP6, and by GIACS (General Integration of the Applications of Complexity in Science).  

\section{Definition}
In this paper, we show how simple logistic growth that was studied intensively during 
the last 200 years in many domains of science could be extended in a rather simple
way. The resulting extended model has, among other features, two very important ones: Intermittency and Localization.
These features were observed repeatedly along the history of science in an enormous number of real-life systems in Economics, Sociology, Biology, Ecology
and more. We suggest by this a unified theoretical "umbrella" that might serve in a surprising way many scientific disciplines who share similar observed patterns.

\section{Introduction}
A well known joke, that many physicists like to tell during their talks in order 
to demonstrate the strength of simplifying the problem one has in hand is: 
"First, let us consider a spherical cow...".
Although, no one really believes in spherical cows - the power of simplification is well
accepted and appreciated by the Physics community, or as Albert Einstein put it, 
very accurately: "Everything should be made as simple as possible, but not simpler".

There are many more such "mantras" like: "Keep it simple, stupid",
"Kill your darlings" and "Less is more". As these lines of thought were adopted so 
strongly by physicist for so much time, the statement of P.W. Anderson that
"More is different" made such a revolution in Science. In the paper that has this
title, Anderson pushed the new scientific (inter -) discipline, now 
known as "complexity". By introducing these new ideas, Anderson paved the 
way for many physicists carrying with them heavy weapons from traditional physics 
to start thinking and attacking many problems from a variety of scientific disciplines.

A lot of criticism about such physicists that try to cross the borders of their
discipline is about \textit{over-simplifying} real-life problems in order to be able to
solve the resulting models with the tools they already have. Due to that, it is important 
to emphasize here that by working inside the framework of complexity one 
tries not to loose the minimal theoretical ingredients of the problem that are sufficient 
to produce the complex observed outcome. Rather than this, one tries 
to study to the best of one's ability, the simplest \textit{possible} model.

A common question that arises in the social sciences is: \textit{Why Improbable Things are so Frequent?}: Fine-tuned irreducibly complex systems have generically a low probability to appear and highly integrated$\simeq$arranged systems are usually "artificial" (often man-made) and untypical. Yet many complex systems are found lately to be "self-organized". More precisely, the amount of non-generic, fine tuned and highly integrated systems is much larger in nature from what would be reasonably expected from generic stochastic estimations. It often happens that even though the range of parameters necessary for some nontrivial collective phenomenon to emerge is very narrow (or even an isolated single "point" out of an continuum infinite range), the phenomenon does actually take place in nature. This leads to collective objects whose properties are not explainable by the generic dynamics of their components. The explanation of the generic emergence of systems which are non-generic from the Multi-Agent point of view seems to be related to self-catalyzing dynamics.

As suggested by the examples above, the frequency with which we encounter non-generic situations in self-catalyzing systems is not so surprising. Consider a space of all possible systems obtainable from certain chemical and physical parts. Even if a macroscopic number of those systems are not auto-catalytic and only a very small number happen to be auto-catalytic after enough time, one of the auto-catalytic systems will eventually arise. Once this happens, the auto-catalytic system will start multiplying leading to a final (or far-future) situation in which those auto-catalytic - a priory very improbable systems - are "over-represented" compared with their "natural" probability of occurrence. Basically, this is how life spread all over Earth.

In this paper, we show how simple logistic growth that was studied intensively during 
the last 200 years in many domains of science could be extended in a rather simple
way and with these extensions is capable to produce a collection of behaviors widely
observed in an enormous number of real-life systems in Economics, Sociology, Biology, Ecology
and more. For other reviews in this direction we recommend on \cite{Stauffer_ea2006,Billari_ea2006,Levy_ea2000} 

The paper will start with a historical overview of the use of logistic-like systems in science 
since its introduction by Malthus in 1798 until today. The next section will present a view of the 
minimal, though sufficient, extensions to the classical logistic system that are able to bring
this theoretical framework, closer to reality, but "auto-catalysis" yet still solvable analytically in many 
regions of the parameter's space. Then, we will show some of the successes we had in applying
this framework to real-life systems. We will finish the paper by a short fantasy trying to describe
a dream about the possible usages of this powerful theoretical framework in the so called 
"soft" sciences in the future.

\section{Logistic systems: From Malthus until today.}

\subsection{"auto-catalysis"}
One of the key concepts underlying the emergence of complex macroscopic features is auto-catalysis. We therefore give at this point a provisory definition of it:
auto-catalysis = self-perpetuation, $\simeq$ reproduction, $\simeq$ multiplication.
As opposed to the usual stochastic systems in which the microscopic dynamics changes typically the
 individual microscopic quantities by additive steps (e.g. a molecule receiving or releasing a quantum
 of energy), the auto-catalytic microscopic dynamics involve multiplicative changes (e.g. the market 
worth of a company changes by a factor (index) after each elementary transaction). Such auto-catalytic 
microscopic rules are widespread in chemistry (under the name of auto-catalysis), biology 
(reproduction $\simeq$ multiplication, species perpetuation), social sciences (profit, returns, rate of growth).

The "autocatalytic" essence of the growth processes was formally
expressed as early as 1798 by T.R. Malthus\cite{Malthus1798} who wrote
a differential equation for describing the dynamics of a population of
proliferating individuals:
\begin{equation}
  \frac{d W (t)}{d t} = a \cdot W (t) \label{eqMAL}
\end{equation}
The growth rate of the population $W$ is proportional to $W$ itself and parametrized by a relative growth (/proliferation) rate $a$ .
The Malthus equation can be reinterpreted to represent a very wide
range of phenomena in various  fields:  behavior adoption in sociology, proliferation in biology, capital returns in economics , or
proselytizing in politics. The (exponential) solution $\sim e^{(a\cdot
t)}$ of this equation influenced much of the subsequent ideas in
various fields and in particular it roused the first worries about the
sustainability of growth. Malthus himself expressed great concern of
the humanitarian "catastrophe" that unlimited population growth may
lead to. However, Verhulst {\cite{Verhulst1838}} introduced (in 1838)
a nonlinear interaction term $-b\cdot W^2$ (that may represent
(confrontation over) limited resources in biology, competition in
economics, limited constituency in politics and finite population in
sociology)
\begin{equation}
  \frac{d W (t)}{d t} = a  \cdot W (t) - b\cdot W^2 (t) \label{eqVER}
\end{equation}

By including this term, rather than increasing indefinitely,  the
solution saturates at a constant asymptotic value $W \longrightarrow
\frac{a}{b}$ . For the following two centuries, this "logistic
dynamics" was considered by the leading scientists as a crucial
element in various fields from biology (Volterra{\cite{Volterra1931}})
to "the everyday world of politics and economics" (Lord
May{\cite{May1976}}).

\subsection{Real-Life examples}                  
\label{Real}
\subsubsection{The A(utocatalysis)-Bomb}
\label{A-Bomb}
The first and the most dramatic example of the macroscopic explosive power of the Multi-agent auto-catalytic 
systems is the nuclear (Atom) bomb. The simple microscopic interaction underlying it is that the U235 nucleus, 
when hit by a neutron splits into a few energetic fragments including neutrons:  
\begin{equation}
 n + U \longrightarrow n + n + etc.
\label{uranium}
\end{equation} 

On the basis of  (autocatalysis equation 1) even without knowing what is a neutron or a U235 nucleus, 
it is clear that a macroscopic “reaction chain” may develop: if there are other U235 nuclei in the neighborhood, 
the neutrons resulting from the first (autocatalysis equation 1) may hit some of them and produce similar new reactions. 
Those reactions will produce more neutrons that will hit more U235 that will produce more neutrons….. 

The result will be a chain (or rather "branching tree") of reactions in which the neutrons resulting from one generation 
of fission events induce a new generation of fission events by hitting new U235 nuclei. This "chain reaction" will go on 
until eventually, the entire available U235 population (of typically some  $10^{26}$ nuclei) is exhausted and their corresponding 
energy is emitted: the atomic explosion. 
The crucial feature in the equation above, which we call "auto-catalysis", is that by inputting one neutron n in the reaction 
one obtains two (or more) neutrons (n+n). 
The theoretical possibility of iterating it and have an exponentially increasing macroscopic number of reactions was explained 
in a letter from Einstein to President Roosevelt.  But only the later attack on Pearl Harbor lead to the initiation of the Manhattan project and the eventual 
construction of the A-bomb.

It is not by chance that the basic Multi-Agent method (the Monte Carlo simulation algorithm used until this very day in physics 
applications) was invented by people (Metropolis, Rosenbluth, 
Rosenbluth, Teller, Teller) involved in nuclear weapons research: the Multi-Agent method is the best fit method to compute realistically the macroscopic 
effects originating in microscopic interactions!

\subsubsection{The B-Bomb: autocatalysis and localization in immunology}
\label{B-Bomb}

In no field is the auto-catalysis and localization more critical than in the emergence of living organisms 
functions out of the elementary interactions of cells and enzymes. 
From the very beginning of an embryo development the problem is how to create a "controlled chain reaction" 
such that each cell (starting with the initial egg) divides into similar cells, yet spatio-temporal structures 
(systems and organs) emerge. Let us consider the immune system as an example. The study of the Immune System
for the past half century has succeeded in characterizing the key: cells, molecules, and genes. As always in 
complex systems, the mere knowledge of the microscopic world is not sufficient (and, on the other hand, some details of
 the micros are not necessary).  Understanding comes from the identification of the relevant microscopic 
interactions and the construction of a Multi-Agent Simulation with which to demonstrate in detail how the complex 
behavior of the immune system emerges. Indeed, the immune system provides an outstanding example of the emergence of 
unexpectedly complex behavior from a relatively limited number of simple components interacting according to 
known simple rules. 
By simulating their interactions in computer experiments that parallel real immunology experiments, one can 
check and validate the various mechanisms for the emergence of collective functions in the immune system. 
(E.g. recognition and destruction of various threatening antigens, the oscillations characteristic to 
rheumatoid arthritis, the localization of diabetes 1 to pancreatic islets etc). This would allow one to design 
further experiments, to predict their outcome and to control the mechanisms responsible for various 
auto-immune diseases and their treatment.

\subsubsection{The Tulip Bomb.}
\label{tulip}
The tulip mania is one of the most celebrated and dramatic economic bubbles in history. It involved the rise of the tulip 
bulb prices in 1637 to the level of average house prices. In the same year, after an increase by a factor of 20 
within a month, the market collapsed back within the next 3 months. After loosing a fortune in a similar event 
(triggered by the South Sea Co.) in 1720 at the London Stock, Sir Isaac Newton was quoted to say,
 "I can calculate the motions of the heavenly bodies, but not the madness of people."

 It might seem over-ambitious to try where Newton has failed but let us not forget that we are 300 years later, 
have big computers and had plenty of additional opportunities to contemplate the madness of people.
One finds that global "macroscopic" (and often "catastrophic") economic phenomena are generated by reasonably 
simple buy and sell "microscopic" operations. Much attention was paid lately to the sales dynamics of marketable 
products. Large amounts of data has been collected describing the propagation and extent of sales of new products, 
yet only lately one started to study the implications of the autocatalytic multi-agent reaction-diffusion formalism 
in describing the underlying “microscopic” process. \cite{Weisbuch_ea2001a,Solomon_ea2000,Goldenberg_ea2000,Yaari_ea2006}

\subsection{Extensions of the classical logistic system.}                  
One of the great early successes of the logistic dynamics was its
application to the spread
of malaria in humans and mosquito's.  Sir Ronald Ross was awarded the
Nobel prize\cite{Ross1911} for this work. His ideas were expressed by
Lotka{\cite{Lotka1923}} in terms of a coupled system of two equations
generalizing \eqref{eqVER}:

\begin{equation} \label{eqLOT}
\begin{array}{cc}
{d w_1 (t)/d t} =a_1\cdot w_1 (t) +a_{12}\cdot w_2 (t)
-a_{112}\cdot w_1 (t)\cdot w_2 (t)
 \\
{d w_2 (t)/d t}=a_2\cdot w_2 (t) +a_{21}\cdot w_1 (t)
-a_{212}\cdot w_1 (t)\cdot w_2 (t)
\end{array}
\end{equation}
Lotka has studied numerically this system in order to predict the
ratios between the infected mosquitoes and the infected humans and the
stability of the system. Vito Volterra advocated independently the use
of equations in biology and social sciences \cite{Volterra1931} and
re-deduced the logistic curve by reducing the Verhulst equation
\eqref{eqVER} to a variational principle that maximed a function that
he named "quantity of life"\cite{Lotka1923}. Later, R.A.
Fisher\cite{Fisher1937} extended of \eqref{eqVER} to spatial
distributed systems and expressed it in terms of partial differential
equations:
\begin{equation}\label{eqFIS}
  \frac{\partial W (\overrightarrow{x},t)}{\partial t} = a \cdot W
(\overrightarrow{x},t) - b\cdot W^2 (\overrightarrow{x},t)
+D\cdot\nabla^2 W (\overrightarrow{x},t)
\end{equation}
He applied this to the spread of a mutant superior gene within a
population and showed that as opposed to usual diffusion, the
propagation consists of a sharp frontier ("Fisher wave") that advances
with constant speed (rather then proportional to $\sqrt{t}$ as in
usual diffusion). Following its formulation, the mathematical study of
{\eqref{eqFIS} was taken over by mathematicians{\cite{Kesten1980}} and
lead eventually a large number of physics studies (especially on the
anomalous and fractal properties of the interface
\cite{BenAvraham_Havlin2000,Grassberger1982,Janssen1981,Cardy_Tauber1996}
).

A crucial step was then taken by Eigen\cite{Eigen1971} and Eigen and
Schuster\cite{Eigen_Schuster1979} who generalized the Lotka system
(3)  of 2 equations for 2 populations to an arbitrary number of
equations $\simeq$populations.
They used the new system in the study of the Darwinian selection and
evolution in prebiotic environments.
More precisely, they considered "quasi-species" of auto-catalytic
(self reproducing RNA sequences) molecules which can undergo
mutations. Each sequence $i$ self-replicates at a rate $a_i$ and
undergoes mutations to other sequences $j$ at rates $a_{ij}$ .
The resulting system of equations is:
\begin{equation}\label{eqEIG}
 \begin{array}{cc}
{d W_i (t)/d t} = a_i W_i (t) + {\sum_{j = 1}^N} a_{ij}  W_j (t)
- {\sum_{j = 1}^N} a_{ji}  W_i (t)- b (  {\overrightarrow{W (t)},
t)})  W_i (t)
\end{array}
\end{equation}

The arbitrary function $b({\overrightarrow{W (t)}, t)}$ represents
generically the interaction with the environment (in the specific case
of ref \cite{Eigen_Schuster1979}  the result of replenishing and
 stirring the container continuously).

The extension of the logistic framework to social sciences was
strongly advanced by Elliot Montroll who based a book on social
dynamics on the principle that "almost all the social phenomena,
except in their relatively brief abnormal times obey the logistic
growth."{\cite{Montroll1978}}.

An analogy that was often exploited in economics was the
ecology-market metaphor (e.g. \cite{Schumpeter1934} ) which was
advanced in parallel with the more mechanical physics analogies. The
connection to the logistic framework was strengthened by the
evolutionary economics metaphor
(e.g.\cite{Nelson_Winter1982,Ebeling_Feistel1982,Jimenez-Montano_Ebeling1980}).
This lead to the extension of \eqref{eqEIG} to economics with the
$a_i$ 's representing capital $\simeq$GDP growth rates and the $a_{ij}$'s
representing trade, social security, mutual help or other mechanisms
of wealth transfer (e.g. taxes $\simeq$ subsidies).
More recently {\cite{Marsili_ea1998}} the logistic dynamics was
applied to the dynamics of the equities $i$ within a personal
portfolio. Then $a_i(t)$'s are interpreted as the rate of growth of
the equity $i$ (at time $t$) and $a_{ij}$ as the periodic
redistribution of capital between the equities by the owner of the
portfolio (in order to optimize it). Stochastic generalizations of the logistic$\simeq$Lotka-Volterra 
equations were studied also in a large body of mathematical literature (e.g. \cite{Kesten1980}), and in order to get meaningful 
results out of the model, one has to introduce the noise in a proper way that will stand for it's effect in real-life systems.

\subsection{The danger of being mean - Simple examples.}
In this sub-section we argue why microscopic (i.e. agent-based) studies are needed
and why simplification in the style of mean field theories can be seriously
wrong. 

If we deal with a small biological population, then due to random accidents 
it may die out completely and irreversibly. For example, poachers may kill the
two surviving males of a small elephant herd which is isolated from other 
elephants. It does not help the herd if one shows that {\it on average} there
is enough food and space for two adult males, two adult females, and several 
calves. For larger populations usually such extreme fluctuations are less
probable, and the time until it happens may increase exponentially with the
population size. 

Also, a hurricane may sink a ship even if averaged over the whole Atlantic
Ocean the absolute value of the wind speed and wave height are moderate. In a 
marriage a husband is supposed to be faithful to his wife and should not average
his efforts to become a father over $10^9$ women; at least that's what wives 
often demand. 

A less trivial example is demography. If you want to know how many people of
retirement age are there for every thousand people of working age, usually 
one takes into account mortalities, birth rates, and migration. Let us assume,
however, that one group of the population has a higher birth rate than the rest
and that this difference is given on to the following generations, either 
genetically or culturally. Then, if everything else is the same, the group with
the higher birth rate will finally dominate in the population, and using the
{\it average} birth rate is not correct. (Of course, if the difference is small
and we want to extrapolate over less than a century, then the average birth
rate is still a good approximation.) One could remedy this error by simulating
the two populations together; but then there could be other inherited traits
which are demographically relevant, and thus with more and finer subdivisions we
finally end up with agent-based demography \cite{bonkowska}, dealing with 
each individual.

This explains the conceptual gap between sciences: in conditions in which only a 
few exceptional individuals dominate, it is impossible to explain the behavior of 
the collective by plausible arguments about the typical or “most probable” individual. 
In fact, in the emergence of nuclei from nucleons, molecules from atoms, DNA from 
simple molecules, humans from apes, there are always the un-typical cases (with 
accidentally exceptional advantageous properties) that carry the day. This effect 
seems to embrace the emergence of complex collective objects in a very wide range 
of disciplines from bacteria to economic enterprises, from emergence of life and 
Darwinism to globalization and sustainability.

In the following section we will bring examples \cite{Shnerb_ea2000} where these 
effects lead to strong localization, such that the mean-field approximations give 
qualitatively wrong results, like predicting extinction where survival is 
possible. The approximations do not become good if only the population is
large.

In conclusion, generic logistic ideas hinted by \eqref{eqVER} arose for the last century in an extremely wide-ranging set of
 applications. For each discipline, subject and system, the variables of the model had to be interpreted in terms of the empirical
 observables and adapted to the relevant range of parameters and initial conditions. Once the parameters are specified, the generic 
framework \eqref{eqEIG} (plus noise) becomes a well defined model for a specific system. Then, one can derive from it precise 
predictions and confront them with the data.

\section{Minimal extensions to the classical logistic system.}

Here, we show how by restricting the parameter's regime of the generic framework \eqref{eqEIG} (plus noise), 
one ends up with a model that has a very strong prediction's power.\cite{Biham_ea1998,Biham_ea2001,Levy_ea2000,Richmond_Solomom2000,Solomon_Richmond2001,Solomon_Richmond2001a,Solomon2000}

\subsection{Case 1: The Generalized Lotka-Volterra System }
If one considers a uniform interaction in \eqref{eqEIG}, the resulting equation can be written as:

\begin{equation}\label{glv}
 \begin{array}{cc}  
{d W_i (t)/d t} = a_i \cdot W_i (t) + \alpha \cdot W (t)- b (  {\overrightarrow{W (t)}, t)}) \cdot W_i (t)
                          \end{array}
\end{equation}
where $W(t)$ is the average value of the $W_i$'s
; then it was shown \cite{Levy_ea2000,Richmond_Solomom2000,Solomon_Richmond2001,Solomon_Richmond2001a,Solomon2000} that :
\begin{itemize}
 \item The system has a steady state for the normalized quantity $X_i(t) \equiv \frac{W_i(t)}{W(t)}$
 \item The steady state distribution of the $X_i$ could be calculated analytically and the resulting distribution has the following form:
$P(X)=e^{{-2\alpha/XD}}\cdot X^{-2-{2\alpha/D}}$ where $D$ is the variance of the distribution from which the growth rates ($a_i$'s) is drawn out of.
 \item The fluctuations of the average ($W(t)$) have a wide distribution with a power-law tail that is closely connected with the value of the steady state distribution (${-2-{2\alpha/D}}$)
\end{itemize}

Obviously, as there is no explicit space in this system, one cannot see localization effects. However, intermittency
is very clear here: The fluctuations of the average value are enormous but changing around a fixed value.
The possible interpretation of such a model are very diverse: 
\begin{itemize}
 \item Income Distribution: $W_i (t)$ can represent the annual income of each individual in the society - then, the $W\alpha$ term is connected to social 
benefits one gets from the being part of the society, such as social security, charity and minimum wage. The $a_i$'s stand for the relative change between 
this year and the previous one. $b (  {\overrightarrow{W (t)}, t)}) \cdot W_i (t)$ then represents the overall trend of the market - periods of depression and of external investments.
 \item Stock Market: $W_i (t)$ can represent the value of a specific stock in the stock market (at the closing time of the market for example) -  then, the $W\alpha$ term is connected to correlations among the different stocks in the market. The $a_i$'s stand again for the relative change between 
the value today and the previous one.  $b (  {\overrightarrow{W (t)}, t)}) \cdot W_i (t)$ represents the overall trend of the market - periods of depression and of external investments.
 \item Population Dynamics: $W_i (t)$ can represent the number of individuals from a specific species in animals or of a specific nation in humans -  then, the $W\alpha$ term is connected to immigration or mutations connecting the different populations.  $b (  {\overrightarrow{W (t)}, t)}) \cdot W_i (t)$ represents the conditions for breeding.
\end{itemize}
There are many more possible interpretations but the point is clear. For each interpretation one can argue that the uniform choice of the interaction matrix is unrealistic - of course it might be true, but as it turns out lately, the power-law prediction is very robust and can stand many different choices of this matrix.
\subsection{Case 2: The "AB Model"}

The "AB Model"\cite{Shnerb_ea2000,Shnerb_ea2001,Louzoun_ea2003a,Louzoun_ea2003,Louzoun_ea2007}} is actually a reaction-diffusion system which has two types of agents: $A$ and $B$. It is a discrete system, both in space and in the fields it describes ($A$ and $B$ in any spatial point are natural numbers, never negative) and as such needs to be described with a set of rate equations. Then the agents may go through the following possible processes with the corresponding rates:

\begin{itemize}
 \item Diffusion: at each time step, with probabilities $D_a/2d$ and 
$D_b/2d$, respectively, an A or B moves to a nearest neighbour site on a 
$d$-dimensional lattice.
\item Reaction: at each time step, with probabilities $\mu$ and 
$\lambda \cdot N_A$, a single $B$ dies or gives birth to a new $B$, respectively, where $N_A$ is the number of $A$'s in the same location.  
\end{itemize}
Naively, this system can be mapped into two partial differential equations:

\begin{equation}\label{eqB}
  \frac{{dB} ( x, t )}{{dt}} = D_b \cdot \nabla^2 B ( x, t ) + (
  \lambda \cdot A ( x, t ) - \mu ) \cdot B ( x, t )
\end{equation}
\begin{equation}\label{eqA}
  \frac{{dA} ( x, t )}{{dt}} = D_a \cdot \nabla^2 A ( x, t )
\end{equation}
It is tempting to say that we can solve equation \eqref{eqA} to get :
\begin{equation}
 {A} ( x, t )\longrightarrow n_A
\end{equation}
in long times and then to plug it into equation \eqref{eqB} to say that depending on the 
parameter $m\equiv(n_A\cdot \lambda - \mu)$ the total number of $B$'s will either increase
exponentially (if $m>0$) or decrease exponentially (if $m<0$). It turns out that this
"mean-field" treatment is totally wrong and as was shown in \cite{Shnerb_ea2000} in low enough
dimensions ($ d \leq 2 $) the $B$'s will asymptotically increase exponentially no matter what
the rest of the parameters are!
The intuitive explanation for this surprising result is that the $B$'s somehow adapt themselves
to be localized around regions with good conditions (large number of $A_i$). One can see a typical
snapshot of this system in Figure \ref{fig1}. Another prediction\cite{Louzoun_ea2003a} of this model is the intermittency of 
the total number of $B$'s even when one adds a saturation term similar to the second term in equation 
\eqref{eqVER}. Yet one more prediction of this model is the "J-shape" in the total number of $B$'s:
i.e. initial decline followed by lasting exponential growth, Figure \ref{fig2}.

\begin{figure}

\resizebox{1\columnwidth}{!}{
  \includegraphics{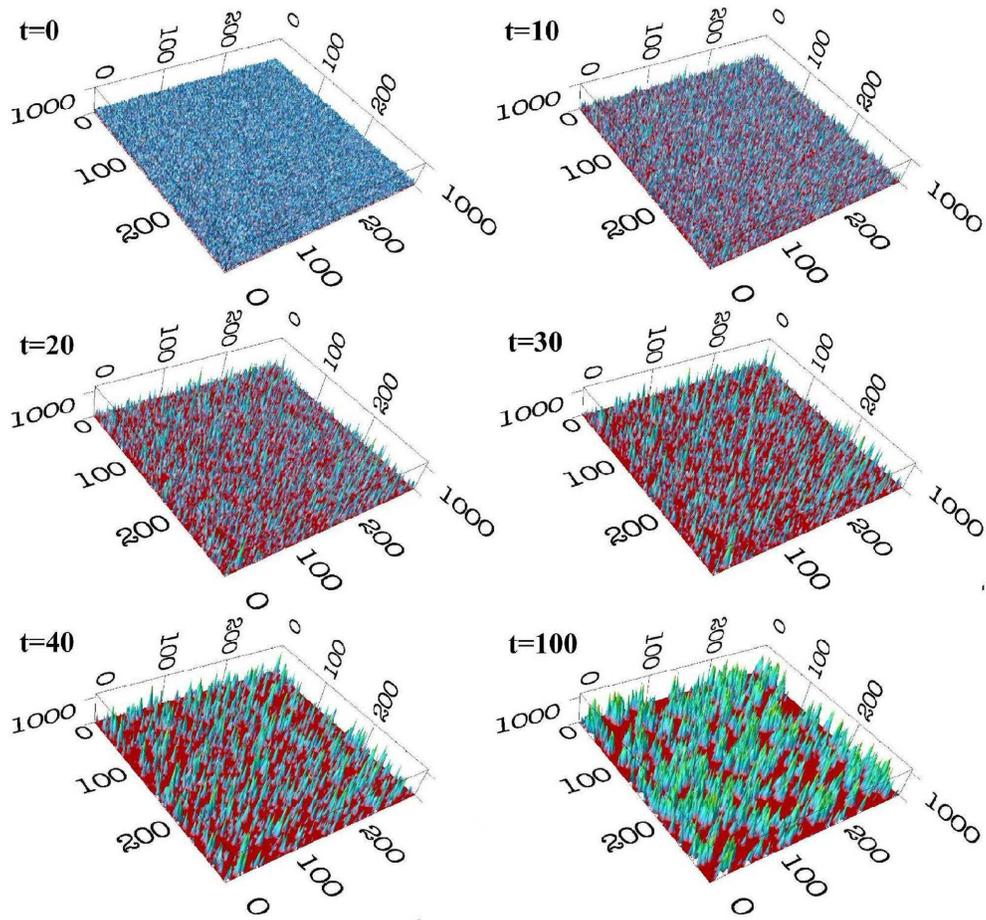}
}
\caption{\textbf{Snapshots of the "AB Model" in two dimensions [Log scale]}\label{fig1}
in these 6 snapshots the time evolution of the "AB Model" is demonstrated:
The two dimension lattice is set in $t=0$ to be in a random distribution of the $B$'s drawn from a Poisson distribution.
The parameters are set in a way that if one looks at the mean-field approximation one will guess that the system needs to go to extinction in a relatively short time. However, due to the discreteness of the catalysts and the reactants, the $B$'s adapt themselves to the rich (in food) areas and the famous "Islands" structure is formed.}

\end{figure}

\begin{figure}

\rotatebox{270}{\resizebox{0.7\columnwidth}{!}{
  \includegraphics{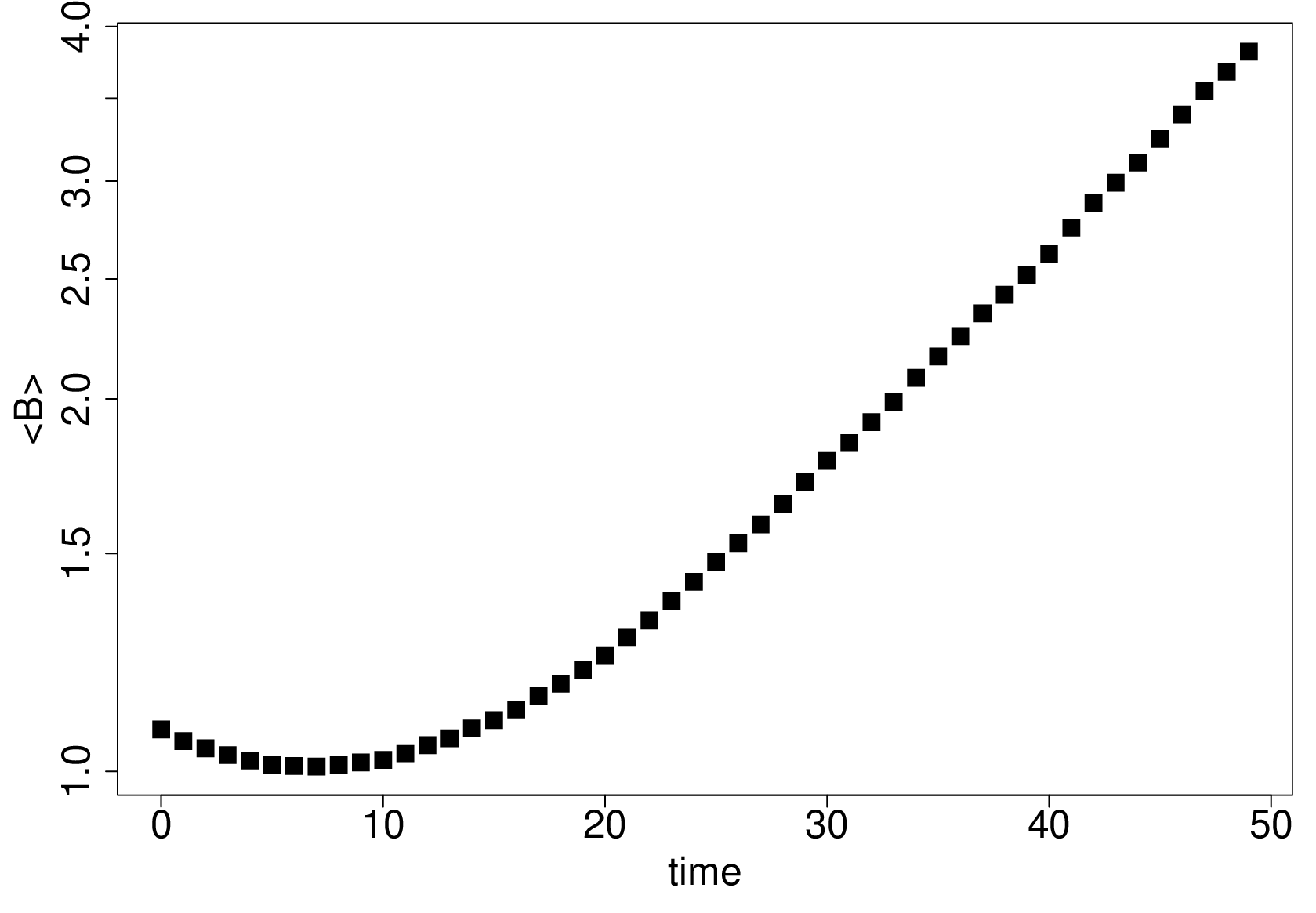}
}}
\caption{\textbf{The time evolution of the average $B$ number in the "AB Model" in two dimensions}\label{fig2}
For the same parameters as Figure \ref{fig1} if one plots the average $B$ number as a function of time, the above picture is revealed: a "J-shape".
It turns out that this shape is a ubiquitous feature for the GDP evolution of many countries that went through a major "shock" in their economies like the collapse of the Soviet bloc for example. }

\end{figure}

\section{Applying these models to real-life systems.}
\begin{figure}

\rotatebox{0}{\resizebox{1\columnwidth}{!}{
  \includegraphics{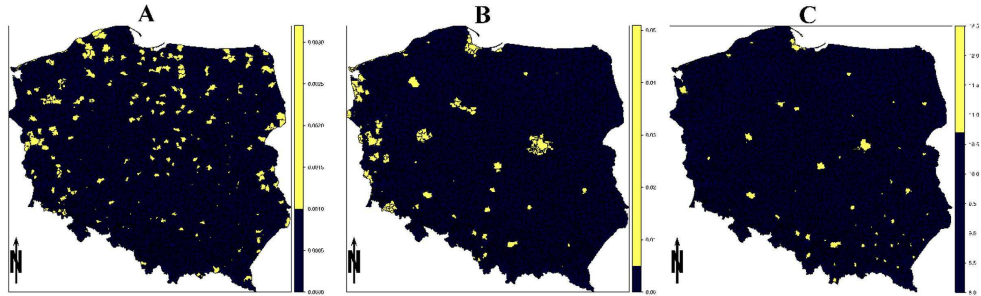}
}}
\caption{\textbf{The influence of Education on Economic Activity before and after
liberalization }
\textbf{ \ref{fig3}A} The number of enterprises per capita in each
county in 1989.
\textbf{ \ref{fig3}B} The number of enterprises per capita in each
county in 1998.
\textbf{ \ref{fig3}C} The years of education per capita in various
counties in 1988.
\textbf{ \ref{fig3}A} maps the number of the enterprises per capita in
the year preceding the economic transition. This initial distribution
does not display any spatial pattern:  is very close to a uniform
random (Poisson) distribution (similar to \ref{fig1} at $t=0$).
\textbf{\ref{fig3}B} After the liberalization there is a clear spatial
pattern: the economic activity is concentrated around the singular
growth centers which are strongly correlated with the education levels
\textit{before} the transition (\textbf{ \ref{fig3}C}). In the
language of the "AB Model", the $A$'s - represent the education level,
while the $B$'s represent the economic activity (enterprises per
capita). \label{fig3}
}

\end{figure}

\begin{figure}

\rotatebox{0}{\resizebox{0.8\columnwidth}{!}{
  \includegraphics{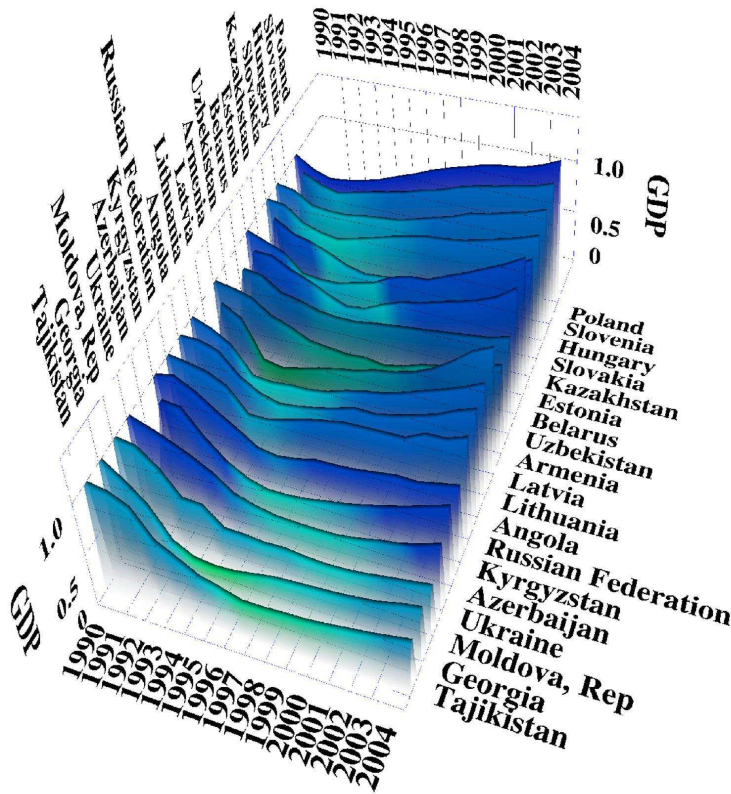}
}}
\caption{ \textbf{The "J-curve" economic recovery after the liberalization of the
Soviet Block} The GDP's of the Eastern European countries experienced
strong decay immediately after the economic liberalization. This was generally followed by a
growth period. The resulting pattern resembles the letter J which
explains the name "J-curve". While the magnitude of the initial decay
and the time and rate of recovery varies among the various countries,
the J-shape is universal. The marked departure (exponential growth) of the Polish economic activity from an exponential decay extrapolated curve indicates that the classical global logistic framework cannot explain the observed pattern. The "AB Model", however does! (as can be seen in figure \ref{fig2})\label{fig4}
}
\end{figure}

As mentioned in \ref{Real}, many real-life systems have characteristics that can be explained with the "AB Model" or the "GLV":\\
In the immune system, it was shown in \cite{Louzoun_ea2001a} that the body's $B$ cells tend to grow in "places" in the genome space where they are needed (depending on the diseases existing in the system). \\
In the Internet, it was shown in \cite{Goldenberg_ea2005} that one can use the theoretical understanding of the model in order to plan strategies that will improve the way the Internet works.\\
In global economics it was shown in \cite{Louzoun_ea2003a} that the global economic system can be mapped into a modification of the  "AB Model". In  \cite{Levy_ea2000} the power law's distribution of the income and of the returns in the stock market were compared to give extraordinary fit with reality. One more paper was just written \cite{Challet_ea2008} on how could decision makers gain knowledge on the economic system they are in charge of under the light of this model. \\
In this section we show in some more details one specific example of 
how one can take the predictions resulting from these types of models 
and apply them to real life systems: The system we will discuss is the Polish economy
following the  collapse of the Soviet bloc. We chose to present the system with the aid of 
equation  \eqref{eqEIG}. In the present application, the index $i$ of the equations in the
system \eqref{eqEIG} ranges from 1 to 2945 and labels the economy of
each of the 2945 counties composing Poland.
Each equation represents the evolution of the economic activity $W_i$
of the county $i$. More precisely, $W_i$ is the number of enterprises
per capita in the county $i$.
The $a_i$'s  represent endogenous growth rate of the county $i$ and
vary from county to county depending on local factors such as social
capital, availability of natural resources or infrastructure. In fact
the data indicate that the most important factor affecting the
economic growth is the education level in the county. This dependence
of a purely economical quantity on a social quantity is of great
methodological importance and emphasizes in a dramatic way the
importance of interdisciplinary studies (in this case economics,
social science and physics). A recent work \cite{Yaari_ea2008} led to a list of nine specific predictions resulting from the model.
The data confirmed in a clear way the model predictions:
Following the liberalization, the counties behaved in divergent ways:
while most of the counties' economies plunged by factors of two, a few
counties tripled their economic activity. This in turn lead to a quick
increase in inequality between the counties. During the preceding
(socialist) regime, all counties were allocated roughly the same
amount of economic activity by the central government. Thus the
counties with high post-liberalization growth rate represented
initially a negligible part of the country GDP and could not avert the
fast global decay. However, within a couple of years, following their
dramatic growth, the fast developing counties became the economic
force, driving up the GDP. Moreover, their influence expanded to the
neighboring regions until, eventually the entire country reached an
uniform growth rate. This is not to say that that the economic
activity per capita equalized. Quite contrary, in the asymptotic
regime in which the growth of the weak regions was due to the
diffusion of economic activity from the fast developing regions, the
very wide differences in GDP per capita persisted and in fact
increased.
One can see in Figure \ref{fig3} the spatial structure of the system, in $t=0$ (year 1989, before the liberalization) and in 1994 and compare it with the social conditions (education level) that catalyzed the economic growth. The localization effect is very clear. In  Figure \ref{fig4} one sees how the generic prediction of the "J-shape" resulting from the "AB Model"is present in all of the formerly communist countries!

\section{Future Directions} 

In his science fiction novel "Foundation" (1951), Isaac Asimov was playing with the idea of having a reliable predictions of the human society under the new scientific discipline he invented and called Psychohistory. In this novel he was dealing mainly with the fact that unlike other scientific disciplines - here, due to the fact that human beings are involved, they are able to read the predictions and by that changing them...Without entering this discussion, we feel that Asimov succeeded to put his finger on a very crucial point: unlike physics for example, there is no such tool that people can rely upon when trying to predict the possible future outcome of today's deeds. Many scientists that come from the natural sciences are very suspicious towards their colleagues from the "soft" sciences because of this reason. On the other hand the "soft" scientists are claiming that the problems that they deal with are far too complex to put into solvable equations. We do understand the positions of both sides, but feel that the time has come to try and close the cultural gap between the two. The methods of the accurate sciences have been improved dramatically since the availability of computer power, on the other hand the social sciences are able today (also due to computers and Internet) to measure many social indexes on a very wide scale and for many years. What is needed now is first to make efforts to quantify the observations that social scientists agree upon, and by that to create a set of so called "stylized facts". After having such list of qualitative and quantitative motifs that science agree they are present in reality - the road to having models that could be validated or invalidated by comparing their theoretical predictions to reality is closer than ever. We do believe that what was described in this paper are first steps towards Asimov's fantasy. Maybe this line of research will help us understand a little better the complex nature of human society.

\bibliographystyle{acm}
\bibliography{poland}

\begin{thebibliography}{10}

\bibitem{BenAvraham_Havlin2000}
{\sc Ben-Avraham, D., and Havlin, S.}
\newblock {\em Diffusion and Reactions in Fractals and Disordered Systems}.
\newblock Cambridge University Press, Cambridge, England, 2000.

\bibitem{Biham_ea2001}
{\sc Biham, O., Huang, Z.~F., Malcai, O., and Solomon, S.}
\newblock Long-time fluctuations in a dynamical model of stock market indices.
\newblock {\em Phys. Rev. E. 64}, 2 (Jul 2001), 026101.

\bibitem{Biham_ea1998}
{\sc Biham, O., Malcai, O., Levy, M., and Solomon, S.}
\newblock Generic emergence of power law distributions and l\'evy-stable
  intermittent fluctuations in discrete logistic systems.
\newblock {\em Phys. Rev. E. 58\/} (1998), 1352--1358.

\bibitem{Billari_ea2006}
{\sc Billari, F., Fent, T., Prskawetz, A., and Scheffran, J.}
\newblock {\em Agent-based computational modelling}.
\newblock Physica-Verlag, Heidelberg, 2006.

\bibitem{bonkowska}
{\sc Bo\'nkowska, K., Szymczak, S., and Cebrat, S.}
\newblock Microscopic modeling the demographic changes.
\newblock {\em Int. J. Mod. Phys. C 17\/} (2006), 1477--1484.

\bibitem{Cardy_Tauber1996}
{\sc Cardy, J.~L., and Tauber, U.}
\newblock On the nonequilibrium phase transition in reaction-diffusion systems
  with an absorbing stationary state.
\newblock {\em {Phys. Rev. Lett.} 77\/} (1981), 4780.

\bibitem{Challet_ea2008}
{\sc Challet, D., Yaari, G., and Solomon, S.}
\newblock The therapy to shock-therapy: optimal dynamical policies for
  transition economies.
\newblock {\em Phys. Rev. E. Submitted\/} (2008).

\bibitem{Ebeling_Feistel1982}
{\sc Ebeling, W., and Feistel, R.}
\newblock {\em Physik der Selbstorganisation und Evolution.}
\newblock Akademie-Verlag, Berlin, 1982.

\bibitem{Eigen1971}
{\sc Eigen, M.}
\newblock Selforganization of matter and the evolution of biological
  macromolecules.
\newblock {\em Naturwissenschaften 58\/} (1971).

\bibitem{Eigen_Schuster1979}
{\sc Eigen, M., and Schuster, P.}
\newblock {\em The Hypercycle}.
\newblock Springer, Berlin Heidelberg New York, 1979.

\bibitem{Fisher1937}
{\sc Fisher, R.~A.}
\newblock The wave of advance of advantageous genes.
\newblock {\em Ann. Eugen. 7\/} (1937), 355--369.

\bibitem{Goldenberg_ea2000}
{\sc Goldenberg, J., Libai, B., Solomon, S., Jan, N., and Stauffer, D.}
\newblock Marketing percolation.
\newblock {\em Physica A Statistical Mechanics and its Applications 284\/}
  (Sept. 2000), 335--347.

\bibitem{Goldenberg_ea2005}
{\sc Goldenberg, J., Shavit, Y., Shir, E., and Solomon, S.}
\newblock Distributive immunization of networks against viruses using the
  'honey-pot' architecture.
\newblock {\em Nature Physics 1\/} (2005), 184 -- 188.

\bibitem{Grassberger1982}
{\sc Grassberger, P.}
\newblock On phase transitions in schlogl's second model.
\newblock {\em Z. Phys. B., Condens. Matter 47\/} (1982), 365--374.

\bibitem{Janssen1981}
{\sc Janssen, H.~K.}
\newblock On the nonequilibrium phase transition in reaction-diffusion systems
  with an absorbing stationary state.
\newblock {\em Z. Phys. B., Condens. Matter 42\/} (1981), 151.

\bibitem{Jimenez-Montano_Ebeling1980}
{\sc Jim\'enez-Montano, M.~A., and Ebeling, W.}
\newblock A stochastic evolutionary model of technological change.
\newblock {\em Collective Phenomena 3\/} (1980), 107--114.

\bibitem{Kesten1980}
{\sc Kesten, H.}
\newblock {\em Random Processes in Random Environments}.
\newblock Springer Verlag, Heidelberg, 1980.

\bibitem{Levy_ea2000}
{\sc Levy, M., Levy, H., and Solomon, S.}
\newblock {\em Microscopic Simulation of Financial Markets: From Investor
  Behavior to Market phenomena.}
\newblock Academic Press, New York, 2000.

\bibitem{Lotka1923}
{\sc Lotka, A.~J.}
\newblock Contribution to the analysis of malaria epidemiology.
\newblock {\em American Journal of Hygiene 3\/} (1923), 1--121.

\bibitem{Louzoun_ea2007}
{\sc Louzoun, Y., Shnerb, N.~M., and Solomon, S.}
\newblock Microscopic noise, adaptation and survival in hostile environments.
\newblock {\em Eur. Phys. J. B 56\/} (2007), 141--148.

\bibitem{Louzoun_ea2001a}
{\sc Louzoun, Y., Solomon, S., Atlan, H., and Cohen, I.~R.}
\newblock The emergence of spatial complexity in the immune system.
\newblock {\em ELSEVIER - Physica A 297\/} (2001), 242--252.
\newblock Received 28 February 2001.

\bibitem{Louzoun_ea2003}
{\sc Louzoun, Y., Solomon, S., Atlan, H., and Cohen, I.~R.}
\newblock Proliferation and competition in decrete biological systems.
\newblock {\em Bulletin of Mathematical Biology 65}, 3 (2003), 375--396.

\bibitem{Louzoun_ea2003a}
{\sc Louzoun, Y., Solomon, S., Goldenberg, J., and Mazursky, D.}
\newblock World-size global markets lead to economic instability.
\newblock {\em Artificial Life 9}, 4 (2003), 357--370.

\bibitem{Malthus1798}
{\sc Malthus, T.~R.}
\newblock {\em An Essay on the Principle of Population}.
\newblock J Johnson, in St Paul’s Churchyard, London, 1798.
\newblock Reprinted by Macmillan and Co, 1894.

\bibitem{Marsili_ea1998}
{\sc Marsili, M., Maslov, S., and Zhang, Y.-C.}
\newblock Dynamical optimization theory of a diversified portfolio.
\newblock {\em Physica A 253\/} (1998), 403--418.

\bibitem{May1976}
{\sc May, R.~M.}
\newblock Simple mathematical models with very complicated dynamics.
\newblock {\em Nature 261\/} (jun 1976), 459--467.

\bibitem{Montroll1978}
{\sc Montroll, E.~W.}
\newblock Social dynamics and the quantifying of social forces.
\newblock {\em Proc. Natl. Acad. Sci. USA 75}, 10 (1978), 4633--4637.

\bibitem{Nelson_Winter1982}
{\sc Nelson, R.~R., and Winter, S.~G.}
\newblock {\em The theory of economic development}.
\newblock Harvard Univ Press, Cambridge, 1982.

\bibitem{Richmond_Solomom2000}
{\sc Richmond, P., and Solomon, S.}
\newblock Power laws are boltzmann laws in disguise, 2000.

\bibitem{Ross1911}
{\sc Ross, R.}
\newblock {\em The prevention of malaria}.
\newblock John Murray, London, 1911.

\bibitem{Schumpeter1934}
{\sc Schumpeter, J.}
\newblock {\em The theory of economic development}.
\newblock Harvard Univ Press, Cambridge, 1934.

\bibitem{Shnerb_ea2001}
{\sc Shnerb, N.~M., Bettelheim, E., Louzoun, Y., Agam, O., and Solomon, S.}
\newblock Adaptatioin of autocatalytic fluctuations to diffusive noise.
\newblock {\em Phys. Rev. E. 63\/} (2001), 21103--21108.

\bibitem{Shnerb_ea2000}
{\sc Shnerb, N.~M., Louzoun, Y., Bettelheim, E., and Solomon, S.}
\newblock The importance of being discrete: Life always wins on the surface.
\newblock {\em Proceedings of the National Academy of Sciences of the United
  States of America 97}, 19 (2000), 10322--10324.

\bibitem{Solomon2000}
{\sc Solomon, S.}
\newblock Generalized lotka volterra (glv) models of stock markets.
\newblock In {\em Applications of Simulation to Social Sciences}. 2000,
  pp.~301--322.

\bibitem{Solomon_Richmond2001}
{\sc Solomon, S., and Richmond, P.}
\newblock Power laws of wealth, market order volumes and market returns.
\newblock {\em Physica A 299}, 1 (2001), 188--197.

\bibitem{Solomon_Richmond2001a}
{\sc Solomon, S., and Richmond, P.}
\newblock Stability of pareto-zipf law in non-stationary economies.
\newblock Computing in Economics and Finance 2001~11, Society for Computational
  Economics, Apr. 2001.
\newblock available at http://ideas.repec.org/p/sce/scecf1/11.html.

\bibitem{Solomon_ea2000}
{\sc Solomon, S., Weisbuch, G., de~Arcangelis, L., Jan, N., and Stauffer, D.}
\newblock Social percolation models.
\newblock {\em Physica A 277}, 1-2 (2000), 239--247.

\bibitem{Stauffer_ea2006}
{\sc Stauffer, D., de~Oliveira, S.~M., and de~Oliveira, P. M.~C.}
\newblock {\em Biology, Sociology, Geology by Computational Physicists}.
\newblock Elsevier, Amsterdam, 2006.

\bibitem{Verhulst1838}
{\sc Verhulst, P.~F.}
\newblock Notice sur la loi que la population suit dans son accroissement.
\newblock {\em Correspondence Mathematique et Physique 10\/} (1838), 113--121.

\bibitem{Volterra1931}
{\sc Volterra, V.}
\newblock {\em Variations and Fluctuations of the Number of individuals in
  animal Species living together}.
\newblock Mc Graw Hill, NY, 1931.

\bibitem{Weisbuch_ea2001a}
{\sc Weisbuch, G., Solomon, S., and Stauffer., D.}
\newblock Economics with heterogeneous interacting agents.
\newblock In {\em Lecture Notes in Economics and Mathematical Systems}.
  Springer, Berlin-Heidelberg, 2001, p.~43.

\bibitem{Yaari_ea2006}
{\sc Yaari, G., Deissenberg, C., and Solomon, S.}
\newblock Advertising , negative word-of-mouth and product acceptance.
\newblock {\em The European Journal of Economic and Social Systems 19}, 2
  (2006).

\bibitem{Yaari_ea2008}
{\sc Yaari, G., Nowak, A., Rakocy, K., and Solomon, S.}
\newblock Microscopic study reveals the singular origins of growth.
\newblock {\em Eur. Phys. J. B Accepted\/} (2008).

\end{thebibliography}
For general books and reviews see refs.4, 18, 38.

\end{document}